\documentclass[lettersize,journal,nosections]{IEEEtran}
\usepackage{amsmath,amsfonts}
\usepackage[linesnumbered,ruled,vlined]{algorithm2e}
\usepackage{array}
\usepackage[caption=false,font=normalsize,labelfont=sf,textfont=sf]{subfig}
\usepackage{textcomp}
\usepackage{stfloats}
\usepackage{url}
\usepackage{verbatim}
\usepackage{graphicx}
\usepackage{cite}
\usepackage{enumitem}
\usepackage{booktabs}
\usepackage{cleveref}
\crefformat{table}{#2Table~\Roman{#1}#3}

\usepackage{threeparttable}
\usepackage{multirow}

\usepackage{fancyhdr} 
\usepackage[switch]{lineno}
\usepackage{xcolor} 

\begin{document}


\title{A Novel HDL Code Generator for Effectively Testing FPGA Logic Synthesis Compilers}

\author{Zhihao Xu, Shikai Guo, Guilin Zhao, Peiyu Zou, Xiaochen Li, He Jiang


\thanks{Z. Xu, S. Guo, and G. Zhao are with the School of Information Science and Technology, Dalian Maritime University, Dalian, China and the Key Laboratory of Artificial Intelligence of Dalian, Dalian, China. E-mail: cemery.xzh@gmail.com, shikai.guo@dlmu.edu.cn, yelen3876@dlmu.edu.cn}

\thanks{P. Zou, X. Li, and H Jiang are with the School of Software, Dalian University of Technology, Dalian, China. E-mail: zoupeiyu@mail.dlut.edu.cn, xiaochen.li@dlut.edu.cn, jianghe@dlut.edu.cn}
}




\maketitle
\begin{abstract}

Field Programmable Gate Array (FPGA) logic synthesis compilers (e.g., Vivado, Iverilog, Yosys, and Quartus) are widely applied in Electronic Design Automation (EDA), such as the development of FPGA programs.
However, defects (i.e., incorrect synthesis) in logic synthesis compilers may lead to unexpected behaviors in target applications, posing security risks. 
Therefore, it is crucial to thoroughly test logic synthesis compilers to eliminate such defects.
Despite several Hardware Design Language (HDL) code generators (e.g., Verismith) have been proposed to find defects in logic synthesis compilers, the effectiveness of these generators is still limited by the simple code generation strategy and the monogeneity of the generated HDL code.
This paper proposes LegoHDL, a novel method to generate syntax valid HDL code for comprehensively testing FPGA logic synthesis compilers.
LegoHDL can generate more complex and diverse defect-trigger HDL code (e.g., Verilog, VHDL, and SystemVerilog) 
by leveraging the guidance of abstract syntax tree and the extensive function block libraries of cyber-physical systems. 
Extensive experiments show that the diversity and defect-trigger capability of HDL code generated by LegoHDL are significantly better than the state-of-the-art method (i.e., Verismith).
In three months, LegoHDL has reported 20 new defects--many of which are deep and important; 16 of them have been confirmed.

\end{abstract}


\section{Introduction}
\label{intro}
Field Programmable Gate Array (FPGA) logic synthesis compilers (e.g., Vivado, Iverilog, Yosys, and Quartus) are indispensable in the field of digital design and engineering, serving as the bridge between theoretical hardware models and implementable digital circuits~\cite{HDL1,yosys,HDL2}. 
Engineers use FPGA logic synthesis compilers to translate high-level descriptions of hardware functionality, expressed in HDLs such as Verilog, into gate-level representations that can be physically implemented on silicon chips. 
This process enables the design of complex Integrated Circuits (ICs), including Central Processing Units (CPUs), Graphics Processing Units (GPUs), and custom Application Specific Integrated Circuits (ASICs).

\begin{figure}[!t]
  \includegraphics[width=\linewidth]{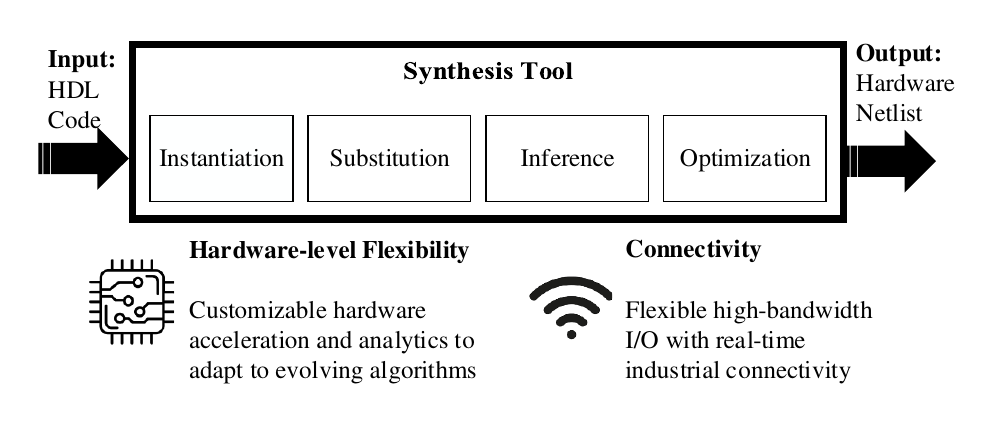}
 \caption{The role of logic synthesis in Intel FPGA playbook\protect\footnotemark}
 \label{fig:1}
 \end{figure}

As shown in Figure~\ref{fig:1}, 
logic synthesis plays a critical role in hardware design development,
which enables different devices (i.e., hardware) to seamlessly integrate with legacy industrial systems. 
This ensures that systems can be promptly updated or reconfigured as needs evolve.
Such adaptability is crucial in industry,
as detailed in Intel's FPGA Industrial Solutions Playbook 2022~\cite{Intels}, 
where it illustrates that FPGA synthesis can lead to more efficient, flexible, and scalable industrial systems. 
However, FPGA logic synthesis compilers are prone to defects since the compilers need to implement intricate optimizations to optimize the power consumption and timing requirements of the designed FPGAs~\cite{verismith}. 
Unexpected behavior of FPGA logic synthesis compilers can impact the final circuit board design and potentially lead to significant losses.

\footnotetext{\url{https://www.intel.com/content/dam/www/central-libraries/us/en/documents/2022-08/intel-fpga-industrial-solutions-playbook-2022.pdf}}



To this end, several methods have been proposed to generate HDL code for testing logic synthesis compilers.
The first method is VlogHammer~\cite{VlogHammer},
which uses a random strategy to generate Verilog code. 
However, VlogHammer~\cite{VlogHammer} cannot generate HDL code containing multiple models; it also does not support behavioral-level Verilog (such as ``always'' blocks).
Verismith~\cite{verismith} is then proposed and becomes the most effective generator for testing FPGA logic synthesis compilers. 
Verismith uses an Abstract Syntax Tree (AST)-based generation method to create HDL code. 
It can generate pseudo-random, valid, deterministic Verilog designs by using the created AST. 
Over a nearly two-year period, Verismith~\cite{verismith} finds 11 defects in logic synthesis compilers.
However, there are still some drawbacks in Verismith~\cite{verismith}, such as the lack of diversity and the high degree of code redundancy for the generated HDL code. 

By a deep analysis of existing methods, we find that two main challenges still remain to be addressed in FPGA logic synthesis compilers testing.
  
\textbf{Challenge 1. Control-flow and data-flow diversity limitation.} 
To uncover deeper, more elusive defects in FPGA logic synthesis compilers, a fundamental requirement is to ensure the adequate complexity of the generated HDL code. 
However, existing HDL code generators, such as Verismith~\cite{verismith}, only employ a simple random strategy to generate HDL code.
The generated HDL code lacks in data-flow and control-flow complexity, thereby hindering the HDL code to comprehensively test FPGA logic synthesis compilers.
Therefore, the first challenge is how to sufficiently generate HDL code with complex and diverse control-flow and data-flow to thoroughly test FPGA logic synthesis compilers.

\textbf{Challenge 2. HDL code type limitation.} 
There are currently three main HDLs commonly used to design and develop ICs or FPGA programs, namely Verilog, SystemVerilog, and VHDL.
However, existing HDL code generators, like Verismith and VlogHammer, predominantly focus on generating Verilog code.
Consequently, the generated HDL code cannot support SystemVerilog and VHDL, which further limits the generality of the generator.
Thus, the second challenge is how to generate HDL code written in different HDLs to test various FPGA logic synthesis compilers.


In this paper, we propose a novel method, named LegoHDL to generate more complex and diverse defect-trigger HDL code for detecting defects in FPGA logic synthesis compilers. 
The basic idea of LegoHDL is to transform the task of HDL code generation into Cyber-Physical System (CPS) model generation.
CPS model is a block-based graphical diagram that can describe the high-level design of hardware systems.
These CPS models can be easily translated into HDL code by CPS model translators (such as HDLCoder in Simulink), and got deployed in the target hardware.

Specifically, LegoHDL has two components: the function generation component and the syntax guidance component. 
Initially, the function generation component randomly connects a set of blocks in CPS function block libraries
to dynamically generate an initial CPS function model (which is similar as a module of a CPS model) from scratch.
The syntax guidance component then runs the HDLCoder tool~\cite{HDLCoder}, which supports to translate the current CPS function model into corresponding HDL code (e.g., Verilog, VHDL, and SystemVerilog), thus addressing the HDL code type limitation challenge.
After that, LegoHDL builds the AST of the generated HDL code. 
Based on the AST, LegoHDL randomly identifies an insertion position in the corresponding CPS function model, and collects the constraint information (e.g., data types and sampling rate) at this insertion position.
LeogHDL then returns the insertion position to the function generation component to guide the subsequent CPS function model generation at this position.
Through dynamic interaction between the AST and CPS function models, LegoHDL can manage the data flow and control flow in HDL code, thereby addressing the control-flow and data-flow diversity limitation. 
The two components continue to interact until the final HDL code and corresponding CPS model are generated, which is similar to Lego by piecing together small pieces of discrete functional component into a complete large project.
Finally, the generated HDL code is used as input to test logic synthesis compilers.

To evaluate the effectiveness of LegoHDL, we ran LegoHDL to generate defect-trigger HDL code on FPGA logic synthesis compilers (i.e., Vivado, Iverilog, Yosys, and Quartus) for three months. 
LegoHDL detected 20 defects, 16 of which have been confirmed by official technical supports.
These defects include those that cannot be detected by state-of-the-art (SOTA) method Verismith~\cite{verismith} (e.g., Yosys defect \#1110 and Quartus defect M82156).


In summary, the main contributions of this work are:

\begin{itemize}
    \item We propose a novel HDL code generator, LegoHDL, which comprises the function generation component and the syntax guidance component to address the control- and data-flow diversity limitation and the HDL code type limitation. 
    
    \item Extensive experiments demonstrate that LegoHDL can generate more defect-trigger HDL code and has identified 20 defects.
   
    \item We have released LegoHDL as a replication package for HDL synthesis compiler testing~\cite{HDLSmith} on GitHub to facilitate future studies.
\end{itemize}

\section{Background and Motivation}

\subsection{The Process of FPGA Logic Synthesis}


\begin{figure*}[!t]
\centering
 \includegraphics[width=0.95\linewidth]{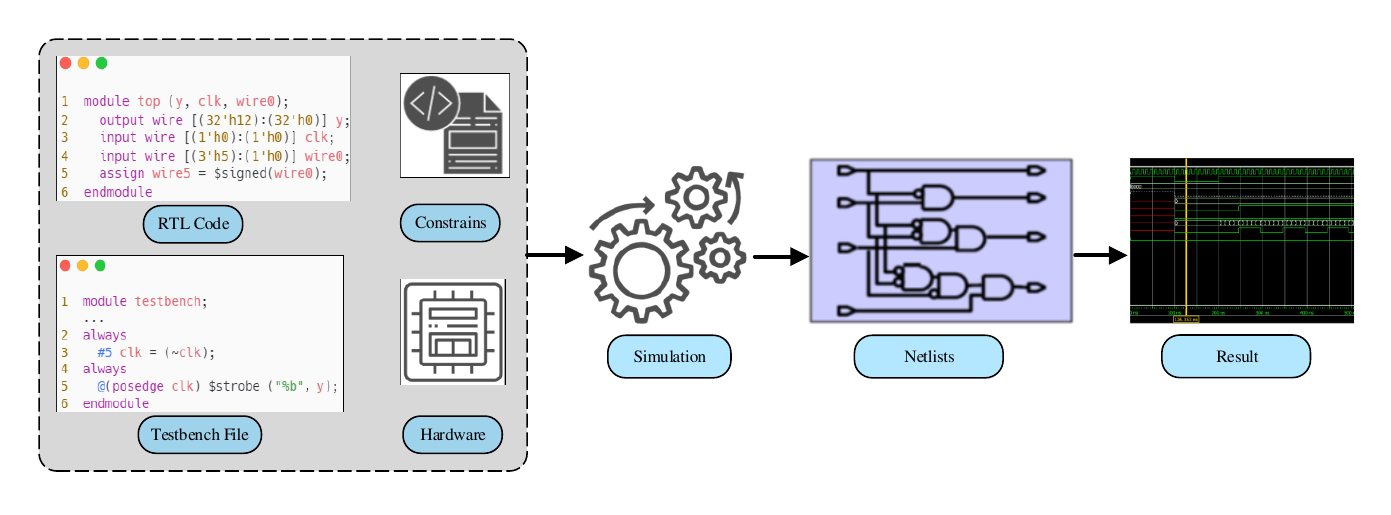}
\caption{The process of logic synthesis}
\label{fig:2}
\end{figure*}


FPGA logic synthesis compilers are integral to FPGA development, as they transform high-level design descriptions into optimized gate-level representations suitable for FPGA hardware. 
These compilers enable designers to focus on the algorithmic and architectural aspects of their designs, automating the conversion of abstract specifications into hardware-implementable logic circuits. 
By optimizing the design for speed, area, and power consumption, FPGA logic synthesis compilers significantly reduce the time-to-market and improve the efficiency of FPGA-based solutions. 
Furthermore, they support iterative design processes, allowing developers to refine their designs based on simulation and real-world testing feedback, ensuring optimal performance and resource utilization in the final product. 
A synthesis process is shown as~\figurename~\ref{fig:2}.

\begin{itemize}
\item The HDL synthesis process commences with the designer authoring the HDL code, which is called Register Transfer Language (RTL) code, detailing the desired hardware functionality and architecture. This RTL code delineates the logical operations, timing, and data flow within the FPGA, ASIC, or SoC. 
\item Upon authoring, the HDL code undergoes verification through behavioral simulation with RTL code and stimulate code, which is called a testbench file. It primarily assesses the design's correctness by comparing it with waveform figures.
\item Once verified via simulation, it progresses through the logic synthesis compiler and output as a netlist. FPGA logic synthesis compilers interpret the HDL code and translate it into a gate-level representation, effectively mapping high-level descriptions to specific hardware elements, such as logic gates, flip-flops, and other foundational digital components.
Simultaneously, the FPGA logic synthesis compilers optimize the design for factors including speed, area, and power consumption, while ensuring compliance with the specified constraints and requirements. They also perform checks to ensure that the design is logically sound and free of errors that could lead to functional failures.
\end{itemize}

After synthesis, the design undergoes place and route, where the physical locations of the logic elements on the FPGA, ASIC or SoC chip are determined and the interconnections are laid out. 
The final output of this process is a bit-stream file, which can be loaded onto the FPGA, configuring its hardware to perform the desired functions as defined by the original HDL code. 

Thus, the synthesis process is crucial, as it bridges the gap between theoretical or conceptual design and the practical, physical implementation in FPGA hardware. If defects exist within the synthesis tool, they could cause unexpected behavior in the final FPGA chip, potentially leading to significant safety hazards.

\subsection{HDL Coder Tool}

HDL Coder\footnote{https://ww2.mathworks.cn/products/hdl-coder.html} is a tool developed by Simulink that allows engineers to automatically generate synthesizable VHDL and Verilog code from CPS models, which could significantly accelerate the design process for FPGA and ASIC implementations by enabling a high-level model-based design approach. 
By utilizing HDL Coder, engineers can focus on algorithm development and system-level design without delving into the intricacies of low-level hardware description languages.
The generated HDL code is optimized for hardware implementation, ensuring efficient utilization of the target FPGA or ASIC resources. 

The process of converting a CPS model into HDL code using HDL Coder involves several steps to ensure the generated code meets the requirements for synthesis and performance. 
Initially, the user designs and simulates the CPS system, where the model's functionality is verified against specifications. 
Following this, HDL Coder performs compatibility checks on the CPS model to identify any constructs that cannot be directly translated into HDL code.

Additionally, HDL Coder offers significant advantages for digital design, notably its ability to generate HDL code in a variety of languages, including Verilog, VHDL, and SystemVerilog. 
This flexibility allows designers to select the language that best suits their project's requirements. 
Furthermore, by leveraging a comprehensive library of CPS function blocks, HDL Coder enables the creation of more complex and sophisticated HDL code designs. 
This capability is crucial for developing advanced digital circuits and systems that require high levels of functionality and performance.

\begin{figure*}[!t] 
\centering
 \includegraphics[width=0.95\linewidth]{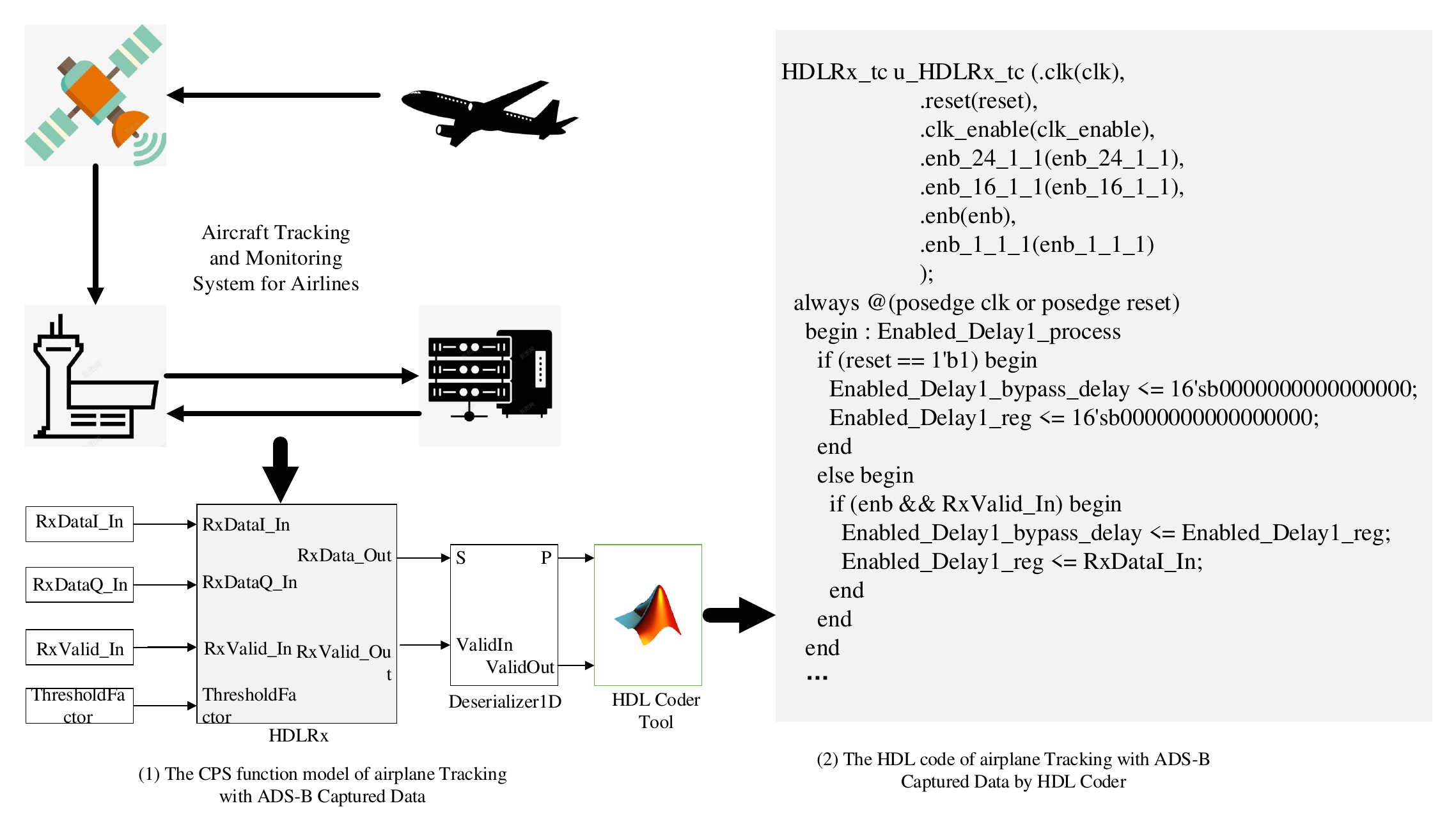}
\caption{Example of translating CPS function model to HDL code by HDL Coder tool}
\label{fig:HDLCoder}
\end{figure*}

The developers may need to modify the model or use HDL-compatible blocks to resolve these issues. 
Subsequently, HDL Coder generates HDL code from the validated model. 
This code generation process includes optimizations such as folding, pipelining, and resource sharing to enhance performance and reduce resource consumption. 
The final step involves verifying the generated HDL code against the original CPS model using HDL simulators or FPGA prototyping, ensuring the hardware implementation accurately reflects the designed system behavior.

As illustrated in~\figurename\ref{fig:HDLCoder}, a CPS function model is translated into multiple lines of HDL function code by HDL Coder. 
\figurename\ref{fig:HDLCoder} (1) depicts a CPS model designed to track airplanes using data captured by Automatic Dependent Surveillance - Broadcast (ADS-B). This model functions as an ADS-B receiver for both HDL code generation and hardware implementation. 
Leveraging the diverse block library of Simulink, LegoHDL provides enhanced capabilities for generating complex HDL code. Additionally, it offers the ability to generate a practical HDL code model.

\section{LegoHDL Model}\label{Model}

In this section, we initially provide a comprehensive overview of LegoHDL. 
Subsequently, we delve into the function generation component and syntax guidance component. 

\subsection{Overview}\label{Overview}

\begin{figure*}[!t]
\centering
  \includegraphics[width=0.85\linewidth]{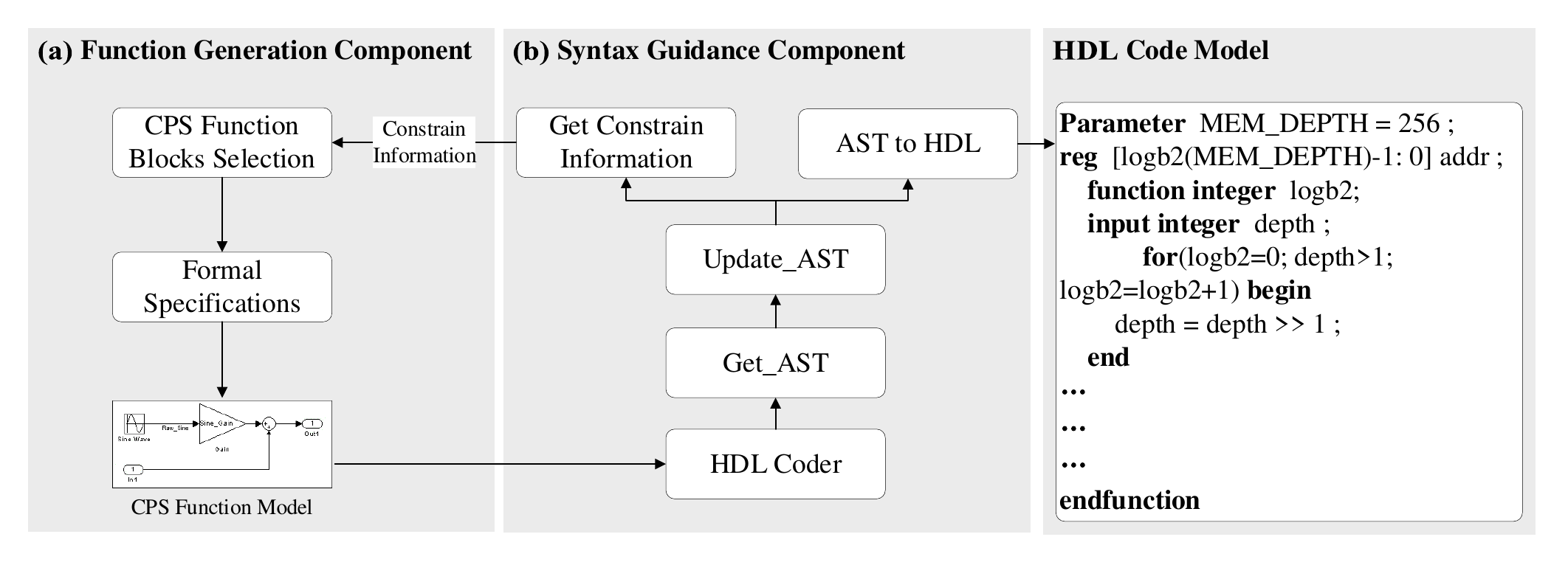}
\caption{The overall framework of LegoHDL}
\label{fig:3}
\end{figure*}

As shown in~\figurename~\ref{fig:3}, LegoHDL has two components: function generation component and syntax guidance component. 
LegoHDL is a fuzzing method designed to generate complex and diverse defect-triggering HDL code for comprehensive testing of logic synthesis compilers, which could help eliminate defects in FPGA logic synthesis compilers. 
Utilizing the extensive model library of CPS models, LegoHDL can generate fully functional HDL code, which differs from Verismith's approach of generating completely random, unreadable HDL code. 
LegoHDL is more in line with the defects encountered by actual engineers during FPGA development.
The process of each component in LegoHDL is delineated in Algorithm \ref{alg:1}. 
Specifically, LegoHDL comprises two components: the function generation component and the syntax guidance component (lines 1-5). 
The function generation component (lines 6-14) primarily selects CPS function blocks from the CPS function block library and interconnects them by the selection of probability matrices to dynamically generate diverse CPS function models, guided by the constrain information of the syntax guidance component.
Next, the syntax guidance component (lines 15-21) converts these models into HDL code (like Verilog, VHDL, and SystemVerilog), integrating this with the existing code to overcome the limitation HDL code type challenge.
After that, LegoHDL extracts the AST of the latest HDL code. 
Based on the AST, LegoHDL selects the next insertion position and feeds back the current constraint information (accessible insertion points, data types, and sampling rate, etc.) to the function generation component to guide the subsequent CPS function model generation.
By dynamically coordinating between the syntax guidance and function generation components, LegoHDL efficiently manages the HDL code's control-flow and data-flow, thus overcoming the limitations related to control and data-flow diversity.
This process, resembling building with LEGO, continues between the function generation component and syntax guidance component until the final HDL code model is constructed (lines 1-5), forming a large, complete system from smaller functional units.
This adaptability is crucial to generate diverse and complex defect-triggering HDL code to thoroughly exercise logic synthesis compilers.

\subsection{Function Generation Component}
\label{generation}

By using function generation component, LegoHDL generates CPS function models layer by layer, incorporates function blocks and connections through the selection of probability matrices and constrain information, and dynamically interacts with the syntax guidance component.
It consists of two sub-steps, namely the \textbf{CPS function block selection} and the \textbf{formal specification}. 

During CPS function block selection, LegoHDL selects CPS function blocks for inclusion in the CPS function block libraries based on a specific probability matrix. 
Throughout this process, LegoHDL selects suitable function blocks for incorporating them into the CPS function model, guided by the AST feedback from the syntax guidance component. 

Throughout the formal specifications, LegoHDL adds CPS function block to the CPS function model and connects them according to the formal specifications and the guidance by the syntax guidance component. 
In this process, LegoHDL ensures the effectiveness and quality of the generated code through formal specifications.

\subsubsection{\textbf{CPS Function Blocks Selection}}

We have collected CPS function block libraries observed in Simulink projects. 
By doing this LegoHDL can better leverages CPS's extensive function block library to create diverse CPS function models and ensure a broad coverage of functionality and design patterns. 

As shown in Table~\ref{tab1}, LegoHDL organizes function blocks into four groups based on functionality. After translation, these four function models can cover the essential functions of HDL: \texttt{Sources and Sinks}, \texttt{Mathematical Operations}, \texttt{HDL-specific Operations}, and \texttt{Control Flow Operations}. 

The \texttt{Source and Sink} library is responsible for providing specific periodic inputs to the CPS model or serves as a termination point to mark the end of the CPS model's output and display the output signal, such as \texttt{Constant}, \texttt{Inport}, \texttt{In Bus Element} and \texttt{From Workspace}. 

The \texttt{Mathematical Operations} library includes common mathematical operations such as \texttt{Abs}, \texttt{Add} and \texttt{Algebraic Constraint}. 

The \texttt{HDL-specific Operations} library provides HDL-specific operations for the model, making the translated HDL code richer and imbued with more hardware design language features, such as \texttt{Bit Clear}, \texttt{Bit Set} and \texttt{Bit to Integer Converter}.

\begin{algorithm}[!t]
\small
\SetAlgoLined
\DontPrintSemicolon
\SetKwInput{KwInput}{Input}
\SetKwInput{KwOutput}{Output}
\SetAlgoNlRelativeSize{-1}
\SetNlSty{textbf}{\color{black}}{}
\SetKwFunction{FModelGeneration}{Function\_Generation}
\SetKwFunction{FSyntax}{Syntax\_Guidance}
\SetKwFunction{FDiff}{Logic\_Synthesis\_Compilers\_Testing}
\SetKwFunction{HDLCoder}{HDL\_Coder}
\SetKwProg{Fn}{Function}{:}{}
\SetKwProg{Pc}{Procedure}{:}{\KwRet}

\KwInput{CPS function model $M$, CPS function block library $B$, Block numbers $N$, HDL Code $H$, AST of HDL code $A$, Constrain information $C$, Generate HDL code iteration number HDL-MAX, Generate CPS function blocks iteration number B-MAX}
\KwOutput{HDL code model}

\For{HDL-MAX}{

    $M \gets$ \FModelGeneration{$B$, $N$, $C$}
   
    $\_, H \gets$ \FSyntax{$M$}

    $\text{HDL code} \gets$ $H$

}

\Fn{\FModelGeneration{$B$, $N$, $C$}}{

    \For{B-MAX}{

    \For{$b \in N$}{
    
        $M' \gets M.\text{CPS\_Function\_block\_Selection}($B$, $C$)$
        
        $M'' \gets M.\text{Formal\_Specifications}($M'$)$
               
    }

    $C$, \_ $\gets$ \FSyntax{$M''$}


    }

    \KwRet{$M'$}
   
}

\Fn{\FSyntax{$M$}}{

    $H \gets \text{HDL\_Coder}(M)$

    $A \gets \text{Get\_AST}(H)$

    $A' \gets \text{Update\_AST}(A)$    

    $C \gets \text{Get\_Constrain\_Information}(A')$    

    $H' \gets \text{AST\_to\_HDL}(A')$    
    
    \KwRet{$C$,$H'$}
    
}

    \KwRet{$\textnormal{HDL code}$}

    
    

    

    
    

    

\caption{LegoHDL model}
\label{alg:1}
\end{algorithm}

The \texttt{Control Flow Operations} library includes Select Model Relationship blocks that can enrich the control flow of the model, such as \texttt{Function-Call Subsystem}, \texttt{If}, and \texttt{If Action Subsystem}. 

Then, LegoHDL assigns predetermined weights to each library, and using these weights, the "choose blocks" method selects blocks to generate CPS models.
Specifically, given a probability matrix \(S\), where each element \(S_{i,j}\) denotes the probability of transitioning from model \(i\) to model \(j\), and a function \(R(m)\) that represents the sampling rate of model \(m\), the following formula ensures the selection of the next model \(m_{\text{next}}\) from the current model \(m_{\text{cur}}\). 
The conditions for selection stipulate that \(m_{\text{next}}\) must be different from \(m_{\text{cur}}\) and must have the same sampling rate.

Equation~\eqref{eq:1} ensures that each potential \(m_{\text{next}}\) has a non-zero transition probability from \(m_{\text{cur}}\), while maintaining that the models are distinct and share the same sampling rate.

\begin{table*}[!t]
  \centering
    \begin{threeparttable}
    \caption{The four-group CPS function block libraries}
      \label{tab1}
    \begin{tabular}{lll}
    \toprule
    \multicolumn{1}{l}{\textbf{Class}} & \textbf{Name}  & \textbf{Function} \\
    \midrule
    \multirow{4}[0]{*}{\textbf{Source and Sink}} & Constant & Generate constant value \\
          & From Workspace & Load signal data from workspace into CPS model \\
          & In Bus Element & Select input from external port \\
          & Inport & Create input port for subsystem or external input \\
    \midrule
    \multirow{15}[0]{*}{\textbf{Mathematical Operations}} & Abs   & Output absolute value of input \\
          & Add   & Add or subtract inputs \\
          & Algebraic Constraint & Constrain input signal \\
          & Assignment & Assign values to specified elements of signal \\
          & Bias  & Add bias to input \\
          & Complex to Real-Imag & Output real and imaginary parts of complex input signal \\
          & Divide & Divide one input by another \\
          & Dot Product & Generate dot product of two vectors \\
          & Find Nonzero Elements & Find nonzero elements in array \\
          & Gain  & Multiply input by constant \\
          & Math Function & Perform mathematical function \\
          & MinMax & Output minimum or maximum input value \\
          & Trigonometric Function & Specified trigonometric function on input \\
          & Unary Minus & Negate input \\
          & Weighted Sample Time Math & Support calculations involving sample time \\
          \midrule
    \multirow{10}[0]{*}{\textbf{HDL Specific Operations}} & Bit Clear & Set specified bit of stored integer to zero \\
          & Bit Set & Set specified bit of stored integer to one \\
          & Bit to Integer Converter & Map vector of bits to corresponding vector of integers \\
          & Bitwise Operator & Specified bitwise operation on inputs \\
          & Combinatorial Logic & Implement truth table \\
          & Compare To Constant & Determine how signal compares to specified constant \\
          & Compare To Zero & Determine how signal compares to zero \\
          & Detect Change & Detect change in signal value \\
          & Detect Decrease & Detect decrease in signal value \\
          & Detect Increase & Detect increase in signal value \\
          \midrule
    \multirow{4}[0]{*}{\textbf{Control-flow Operations}} & Function-Call Subsystem & Subsystem whose execution is controlled by external function-call input \\
          & If    & Select subsystem execution using logic similar to if-else statement \\
          & If Action Subsystem & Subsystem whose execution is enabled by If block \\
          & Model & Reference another model to create model hierarchy \\
           \bottomrule
    \end{tabular}%
        \end{threeparttable}
\end{table*}%

\begin{equation}\label{eq:1}
\begin{split}
\forall m_{\text{next}} \in M &: S_{m_{\text{cur}}, m_{\text{next}}} > 0 \Rightarrow \\
&\left( m_{\text{next}} \neq m_{\text{cur}} \wedge R(m_{\text{next}}) = R(m_{\text{cur}}) \right)
\end{split}
\end{equation}

\subsubsection{\textbf{Formal Specifications}}

After selecting the appropriate CPS function block $b$ from the CPS function libraries $B$, LegoHDL uses it as its child function block.
LegoHDL is required to satisfy two formal specifications (i.e., data type formal specification and sampling rate formal specification) as illustrated in Equation~\eqref{eq:formal}, which could improve the success rate of CPS function model generation.
The data type formal specification $C_{S} (m)$ must belong to the range of data types $W$ supported by the CPS function model, as returned by the syntax guidance component for the generated CPS function model information.
The sampling rate formal specification needs to be adjustable to align with the requirements of the generated CPS function model $C_ {b} (b)$, eliminating any external interference. Additionally, \(\lambda(m) = 0\) ensures no model partakes in dependency loops; it is crucial for HDL generation. 
Because Combinational Loop violate the principle of synchronization design can easily produce oscillations, glitches and timing violations, making the entire system extremely unstable~\cite{combinational-loop}.

 \begin{equation}
  \forall  b  \in  BL: C_ {S}  (m)  \in  W  
  \wedge AST(C_ {b}  (b),b) \wedge 
  \lambda(m) = 0 
   \label{eq:formal}
 \end{equation}

Furthermore, it is essential to impose formal specifications on timing optimization to ensure that the model we generate achieves the shortest possible timing path. 
We have adopted two distinct methods, originating from the CPS function model and the HDL model, respectively. 
Initially, in the CPS function model, we minimize HDL code latency by simplifying model computations and eliminate redundant code. 
Subsequently, in the HDL model, we identify the critical path using AST and timing calculation formulas. 
The critical path represents the longest delay from input to output within a digital circuit. 
In the design of high-performance circuits and systems, the critical path dictates the maximum operating frequency. 
The mathematical representation of the critical path delay in digital circuits is expressed as,

\begin{equation}\label{eq:critial path}
 D_{\text{critical}} = \max(d(p_1), d(p_2), \ldots, d(p_n))
\end{equation}
where \( D_{\text{critical}} \) denotes the critical path delay. 
The function \( d(p_i) \) calculates the total propagation delay for each path \( p_i \) from an input to an output in the circuit. 
The \( \max \) function is employed to identify the maximum value among all the calculated path delays. 
This maximum delay, which is the longest delay, sets the upper limit on the operational speed of the circuit and determines its maximum clock frequency.
By identifying and optimizing critical paths, we effectively reduce latency and enhance the quality of generated HDL code. Furthermore, it can help us comprehensively test synthesis compilers. 
Finally, throughout the HDL Coder generation process, we explore various optimization levels, compare them, and determine the HDL code with the optimal timing relationship, employing strategies like pipelining optimization~\cite{Pipeline-Optimization}.

In addition, LegoHDL can improve the conversion efficiency of CPS functional model to HDL code and the quality of generated HDL code by combining different optimization strategies.
Equation~\eqref{eq:formal_optimization_comparison} determines the process of optimal optimization strategy choosing. 
\(P(L)\) represents the performance metric for legacy mode, and \(P(i)\) denotes the performance at level \(i\), where \(i\) ranges from 0 to \(n\). 
The function \(\arg\max\) is utilized to select the strategy that maximizes the performance, comparing the legacy mode with each level of the new optimization strategies. 
The final output, \(\text{O}_{\text{best}}\), indicates the strategy with the highest performance.

\begin{equation}
  \text{O}_{\text{best}} = \arg\max \left( P(L), \max_{i=0}^{n} P(i) \right)
  \label{eq:formal_optimization_comparison}
\end{equation}

Through these formal specifications, the generated CPS function models by LegoHDL have higher success rate and better test performance.

\begin{figure*}[!t]
 \includegraphics[width=0.95\linewidth]{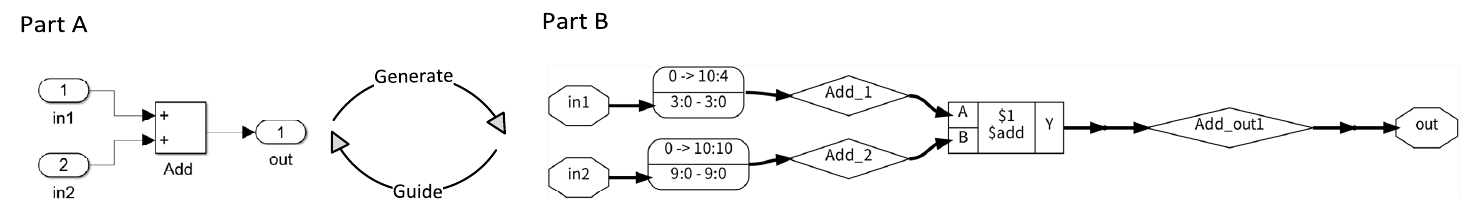}
\caption{Example iteration process of CPS model and AST of HDL Code}
\label{fig:AST}
\end{figure*}

\subsection{Syntax Guidance Component}

The syntax guidance component transforms the CPS functional model into the corresponding HDL code and adds it to the existing HDL code model, which is similar to LEGO by piecing together small pieces of discrete functional component into a complete large project.
Then, we perform AST builds on the updated HDL code model to parse the logical relationships and hierarchies in the code, which lays the groundwork for further analysis and optimization by LegoHDL.
After extracting the AST, LegoHDL will select the appropriate next-level insertion point and feedback the constrain information to the function generation component.
Specifically, we are guided by follow three principles when selecting insertion points.

\textbf{Data Dependency Rule.} 
As illustrated in Equation~\eqref{eq:data dependency}, this rule states that for any nodes \(n\) and \(m\) and any variable \(x\), if \(x\) is used at node \(m\), then \(x\) must have been defined at some node \(n\) where \(n\) is less than \(m\). 
Here, \(D(n, x)\) indicates that \(x\) is defined at node \(n\), and \(U(m, x)\) indicates that \(x\) is used at node \(m\).

\begin{equation}\label{eq:data dependency}
\forall n, m, x : U(m, x) \Rightarrow (D(n, x) \land n < m)
\end{equation}

\textbf{Control Flow Rule.} 
As shown as Equation~\eqref{eq:control flow}, this rule ensures that if node \(n\) controls node \(m\), there should not be another node \(p\) (where \(p \neq m\)) that is also controlled by \(n\) under the same control construct. 
This prevents multiple entries or exits in control structures like loops or conditional branches, which could lead to logical conflicts or errors in execution flow.

\begin{equation}\label{eq:control flow}
\forall n, m : C(n, m) \Rightarrow \neg \exists p : (p \neq m \land C(n, p))
\end{equation}

\textbf{Resource Conflict Avoidance Rule.}
According to Equation~\eqref{eq:Resource Conflict}, this rule ensures if resource \(r\) is used at node \(n\), no other node \(m\) that executes concurrently should use the same resource \(r\). 
It can prevents resource conflicts which can lead to undefined behaviors or errors in the HDL circuit operation.

\begin{equation}\label{eq:Resource Conflict}
\forall n, m, r : (R(n, r) \land R(m, r)) \Rightarrow n = m
\end{equation}

Armed with the selecting insertion points, LegoHDL proceeds to inform the function generation component, providing it with all the necessary parameters (like datatype) and guidelines to construct the next level of the CPS model. 
Thus, the accuracy of the CPS functional model generated by the function generation component can be improved by the information guidance of the syntax guidance component

As demonstrated in~\figurename~\ref{fig:AST}, utilizing the interaction with AST, we have generated a CPS function model. 
This model includes a \texttt{Source} function block, an \texttt{Addition} function block and a \texttt{Display} function block. 
This example has illustrated the interaction of CPS function generation component and syntax guidance component. 
Firstly, we have added two source blocks. 
Their data types are respectively ufix4 and ufix10. 
When translating to HDL code, the width of signal is $3:0$ and $9:0$. 
The syntax guidance component initially determines the data type as ufix based on the data dependency rule and identifies the data bit width as exceeding $9:0$.
Subsequently, the syntax guidance component analyzes the control flow of the entire AST to identify potential insertion points according to the control flow rule. 
Ultimately, in accordance with the resource conflict avoidance rule, the syntax guidance component merges the two input operations into a single signal and relays the data type and insertion point as constrain information back to the function generation component. 
The function generation component selects a block based on the information and probability matrix supplied by the syntax guidance component, then inserts this block into the CPS model adhering to the formal specifications. 

This process is repeated until the \texttt{Source} block is generated.
When \texttt{Source} block has been added in CPS model, LegoHDL will translate the model into HDL code by using HDL Coder. LegoHDL breaks through the limitation of a single generation language when translating CPS model to HDL Code. 
Through HDLCoder we can specify the generated language as Verilog, VHDL or SystemVerilog. 
This approach also solves the problem that previous methods~\cite{VlogHammer,verismith} can only generate single HDL design language.

This iterative process between the syntax guidance and function generation components facilitates a dynamic and responsive design workflow. 
By continuously analyzing and adjusting based on the insight of AST, LegoHDL can adapt to complex design requirements and constraints. 
This adaptability is crucial to generate diverse and complex defect-triggering HDL code to thoroughly exercise the logic synthesis compilers.

\section{Evaluation}

\subsection{Evaluation Setup}\label{Diff}

\begin{figure}[!t]
\centering
\includegraphics[width=0.95\linewidth]{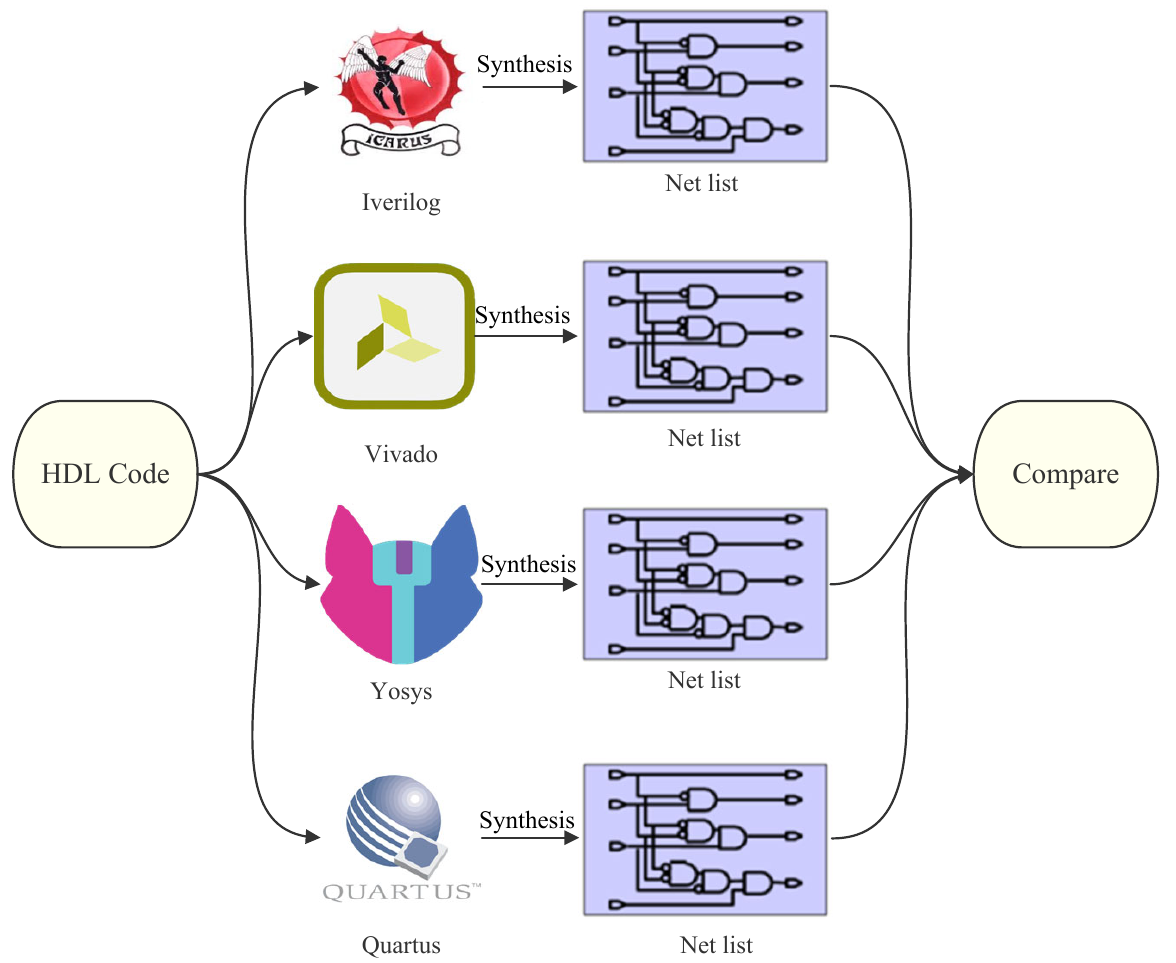}
\caption{Process of Differential Testing}
\label{fig:5}
\end{figure}

For testing FPGA logic synthesis compilers, it is important to determine whether the output results of generation HDL code is correct.
To address these problem, LegoHDL use differential testing apprach to detect defects in logic synthesis compilers, which is shown as \figurename~\ref{fig:5}. 
Differential testing executes the same generation HDL code model in the different synthesis compilers (e.g., Vivado, Iverilog, Yosys, and Quartus) and compares their output results. 
If the outputs differ, it may indicate bugs during the execution of these logic synthesis compilers. 

For Iverilog, Vivado, and Quartus, which support simulation verification functions, LegoHDL has designed testbench files that incorporate CPS simulation result verification and functional logic verification. 
The output results of Iverilog, Vivado, and Quartus are then compared through the simulation verification of stimulus files. 
As Yosys does not support simulation, the equivalence checking tool Sby\footnote{https://github.com/YosysHQ/sby} is employed to verify the functional consistency between the source HDL file and the Yosys-generated output netlist file through formal inspection.

In order to clearly illustrate the root cause of defects, we utilize automated reduction methods to simplify the HDL code that triggers the defect. 
This process enables developers to quickly comprehend and rectify defects. 
Specifically, we employ a technique resembling the binary search method. 
Leveraging the AST extracted from the HDL code, we iteratively delete portions of the code until the erroneous use case can no longer be minimized.
To avoid reporting duplicate defects, we manually use failed assertions and back-trace to detect duplicates. 
When two defects have same failed assertion or back-trace, we consider them as duplicates. 
Finally, we report the detected defects which are verified as non-duplicates as new issues to official technical supports (e.g., Vivado, Iverilog, Yosys, and Quartus). 

The official websites of various FPGA logic synthesis compilers eventually categorizes each issue into \texttt{Confirmed as New}, \texttt{Confirmed as Known}, Unconfirmed in Doubt (\texttt{Pending Verification}) and not a defect. 
\texttt{New} denotes issues that have been acknowledged as defects previously unknown to developers. 
\texttt{Known} refers to issues recognized as defects that developers were already Known. 
\texttt{Pending} indicates that developers consider the issues avoidable through specific standardized operations.
To facilitate the reproduction of our findings, we have made available the HDL code files that trigger these defects on GitHub\cite{HDLSmith}.

LegoHDL has been developed using MATLAB and Python, and both our code and experimental data are publicly accessible on GitHub~\cite{HDLSmith}. 
The evaluation of LegoHDL was conducted on a computer running the Ubuntu 22.04 operating system, equipped with an Intel Core i9 CPU @ 2.10GHz, and 128GB of memory.

\subsection{Research Questions}

In this section four experiments are conducted to evaluate the efficiencies of LegoHDL. 
Specifically our evaluation aims at answering five Research Questions (RQs).

\textbf{RQ1:} How is the defect-finding capability of LegoHDL in HDL synthesis compilers?

\textbf{RQ2:} Can LegoHDL detect more FPGA logic synthesis compilers' defects compared to the state-of-the-art methods? 

\textbf{RQ3:} Can LegoHDL improve the complexity of HDL code compared to the state-of-the art methods?

\textbf{RQ4:} How efficient is LegoHDL in generating HDLCode?

In our experiments, RQ1 and RQ2 are used to evaluate the defect-finding capability of LegoHDL compared to the state-of-the-art methods. 
RQ3 and RQ4 are employed to evaluate AST generation for improving the complexity and improving success generation rates.

\subsection{State of The Art Method}

Since the VerilogHammer~\cite{VlogHammer} version is too old and does not support the latest HDL syntax rules, we did not choose it as our baseline method.
We choose Verismith~\cite{verismith} as our state of the art method, as only Verismth is the fuzzer tool of Verilog coder in FPGA logic synthesis compilers test. 
We reproduce Verismith~\cite{verismith} with source code provided by their works and use their default configurations.

\begin{table*}[!t]
  \centering
  \caption{Defects found by LegoHDL}
  \begin{threeparttable}
    \begin{tabular}{llllll}
    \toprule
    Num & ID    & Summary & Status & Type  & Software \\
    \midrule
    1&\#4279 & Top package import error in Systemverilog & Verified & C     & Yosys \\
    2&\#4276 & Bit-width type conversion causes array overflow & Verified & C     & Yosys \\
    3&\#4277 & latch inferred in synthesis process by complexity reference relationship & Verified & C     & Yosys \\
    4&\#4217 & Syntax error: assign a variable at the same time & Pending & M     & Yosys \\
    5&\#1110 & SystemVerilog reference error & Verified & M     & Iverilog \\
    6&EMtSSSA1 & Synthesis crash under unsigned delay of specific symbol & Verified & C     & Vivado \\
    7&WR8MSAW & Synthesis faild and crush(HARTRegInfo::Synchronousity) & Verified & C     & Vivado \\
    8&EOHpUSAX & Vivado synthesis faild (libc.so.6 error) & Verified & C     & Vivado \\
    9&EOHuZSAX & Vivado synthesis failed (librdi\_synth.so) & Verified & C     & Vivado \\
    10&GOZ7SAO & Synthesis failed caused by groupMFFC  function. & Verified & C     & Vivado \\
    11&H5C2SAK & Synthesis failed caused by callABC  function. & Verified & C     & Vivado \\
    12&H19BSAS & Reference relationship error casued synthesis failed & Pending & C     & Vivado \\
    13&FXlJSAQ & Add component overflow & Pending & M     & Vivado \\
    14&C1GJxSAN & Global RAM value reset inconsistencies & Pending & M     & Vivado \\
    15&DL0yBSAT & Synthesis failed caused by hdi::tcltasks::task\_manager::eval\_in\_tcl & Verified & C     & Vivado \\
    16&E6QSDSA3 & Synthesis failed caused by ConstProp::assertPortEquivalencies & Verified & C     & Vivado \\
    17&E6QUESA3 & Synthesis failed caused by HARTRegInfo::isValidForBlock & Verified & C     & Vivado \\
    18&E6QudSAF & Synthesis failed caused by NNetC::singleDriver & Verified & C     & Vivado \\
    19&M82000 & Synthesis crashed caused by code 0x3f0e5 & Verified & C     & Quartus \\
    20&M82156 & Arry value passed over 2 bits & Verified & M     & Quartus \\
    \bottomrule
    \end{tabular}%
     \begin{tablenotes}
        \footnotesize
        \item[1] There are two types of status feedback from MathWorks on defect report (i.e., $Verified$ = newly confirmed defect, $pending$ = Pending verification). There are two types of defects (i.e., Type) in our reported defects: crash defects ($C$) and miscompilation defects ($M$).
    \end{tablenotes}
        \label{tabledefect}
    \end{threeparttable}
\end{table*}%

\subsection{Answer to RQ1: Can LegoHDL have the capability of defect-detect in HDL synthesis compilers?}

As demonstrated in Table~\ref{tabledefect}, LegoHDL identified 20 defects over a 3-month period. 
16 defects have been confirmed by official technical supports, and 10 defects have been fixed (e.g., C1GJxSAN, fxlJSAQ, M82000), while 4 defects are scheduled for correction in the next software version. 

\begin{figure}[!t]
  \centering
  \includegraphics[width=0.85\linewidth]{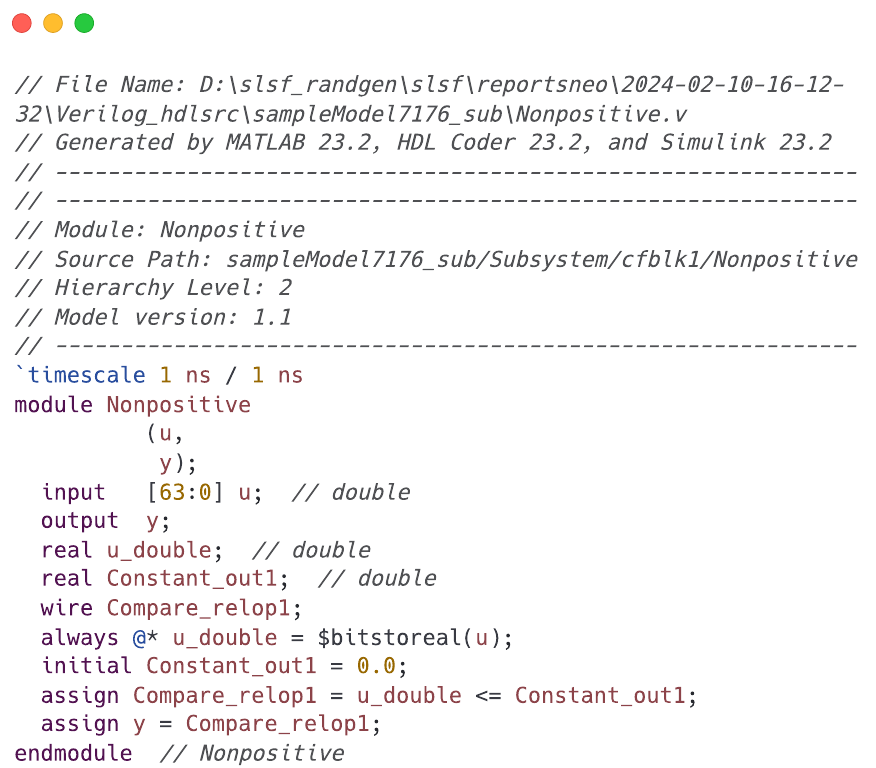}
  \begin{center}
      \footnotesize (a) Reduced code of Defect H19BSAS
  \end{center}
  \includegraphics[width=0.85\linewidth]{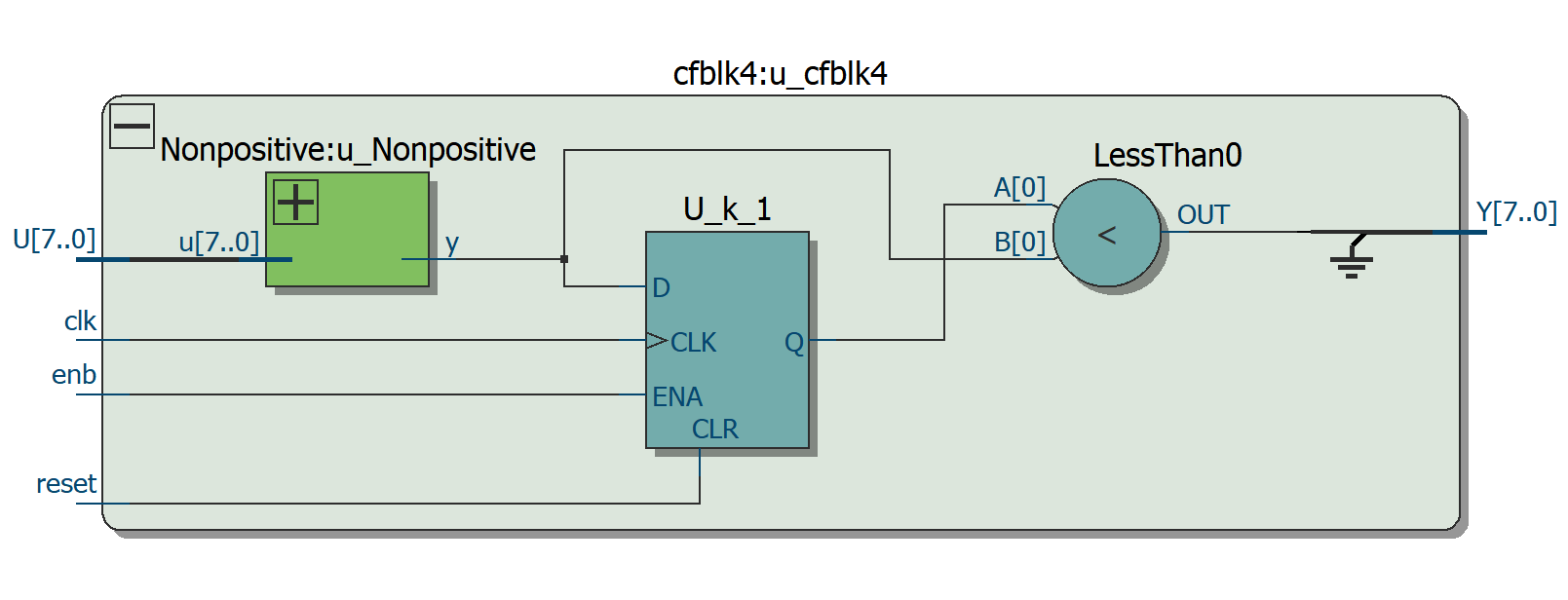}
  \begin{center}
      \footnotesize (b) Netlist of Defect H19BSAS
  \end{center}
  \caption{Reduced Example of Defect H19BSAS}
  \label{fig:6}
\end{figure}

\textbf{Crash of synthesis (Defect H19BSAS).} 
As shown in~\figurename~\ref{fig:6}, we display a defect discovered by LegoHDL. 
Specifically, when synthesizing HDL code, the compiler should correctly handle the reference relationships. 
However, the compiler did not execute the code in the model Nonpositive correctly. 
Nonpositive is a CPS function block that has been correctly transformed to Verilog. 
This complicated reference relationship can cause synthesis errors and potentially lead to compiler crashes. 
Verismith and current methods cannot find this defect because they lack a sufficiently comprehensive corpus to generate code, and they cannot generate complicated reference relationships between blocks. 
Vivado~\cite{Vivado_d} will be affected by this defect.

\begin{figure}[!t]
  \centering
  \includegraphics[width=0.95\linewidth]{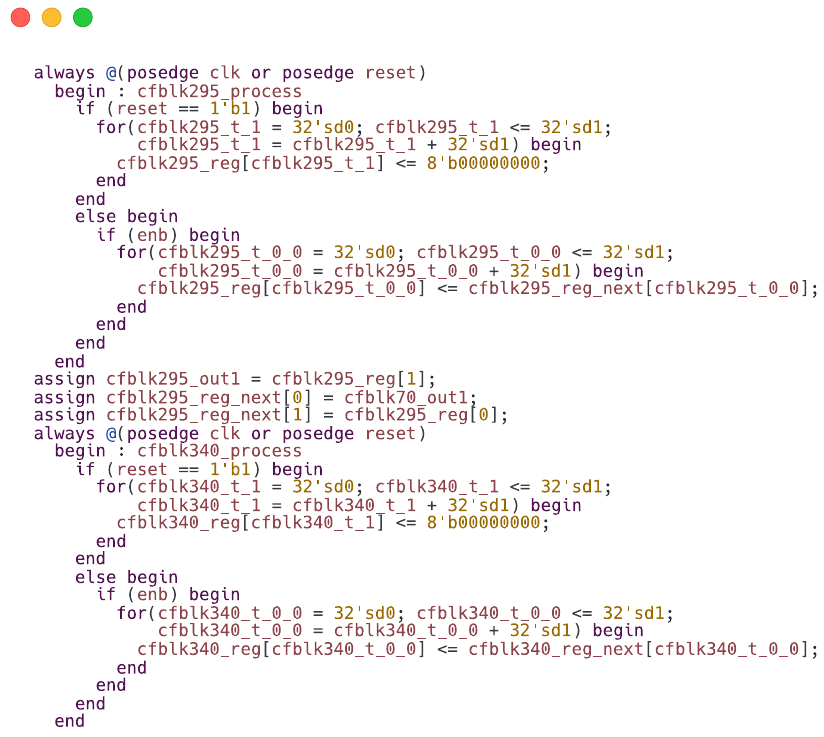}
  \begin{center}
      \footnotesize (a) Reduced code of Defect M82156
  \end{center}
  \includegraphics[width=0.95\linewidth]{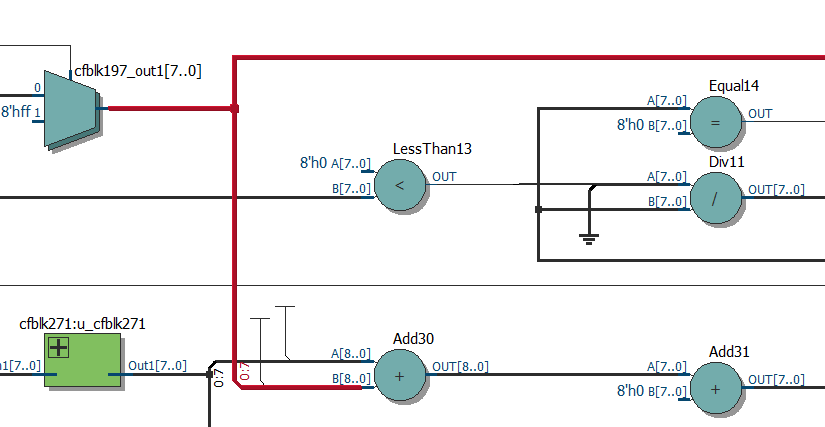}
  \begin{center}
      \footnotesize (b) Netlist of Defect M82156
  \end{center}
  \caption{Reduced Example of Defect M82156}
  \label{fig:7}
\end{figure}

\textbf{Misscompilation of synthesis (Defect M82156).} 
As shown in~\figurename \ref{fig:7}, we present another defect discovered by LegoHDL. 
Specifically, when synthesizing HDL code, the compiler should correctly handle the reset relationship and pass the correct values. 
However, the compiler failed to execute the code during the reset process and caused array overflow of add block (\emph{Add39} in~\figurename \ref{fig:7}), assigning value \emph{XX} from 4000\emph{ps} to 12000\emph{ps}. 
Verismith and current methods cannot identify this defect because they are unable to generate HDL code with complex reset processes and detailed timing information. 
Both Vivado~\cite{Vivado_d} and Quartus~\cite{Quartus} will be affected by this defect. 
It is worth mentioning that the original file, in which the defect was discovered, was written in SystemVerilog. These defects can not be detected by SOTA method Verismith, because it can not generate HDL code written by SystemVerilog code. 

\subsection{Answer to RQ2: Can LegoHDL detect more FPGA logic synthesis compilers' defects compared to the state-of-the-art methods? }

\textbf{Approach.} 
To evaluate the effectiveness of LegoHDL, we compare the defects-finding capability of LegoHDL with the state-of-the-art methods Verismith~\cite{verismith}, since finding more defects within a time period is the main objective of these methods. 
In this experiment we detect defects on the recently released FPGA logic synthesis compilers version. 
We set a single testing period of two weeks for each FPGA synthesis tool, that is every method test FPGA logic synthesis compilers for two weeks.

\begin{figure}[!t]
\centering
 \includegraphics[width=0.95\linewidth]{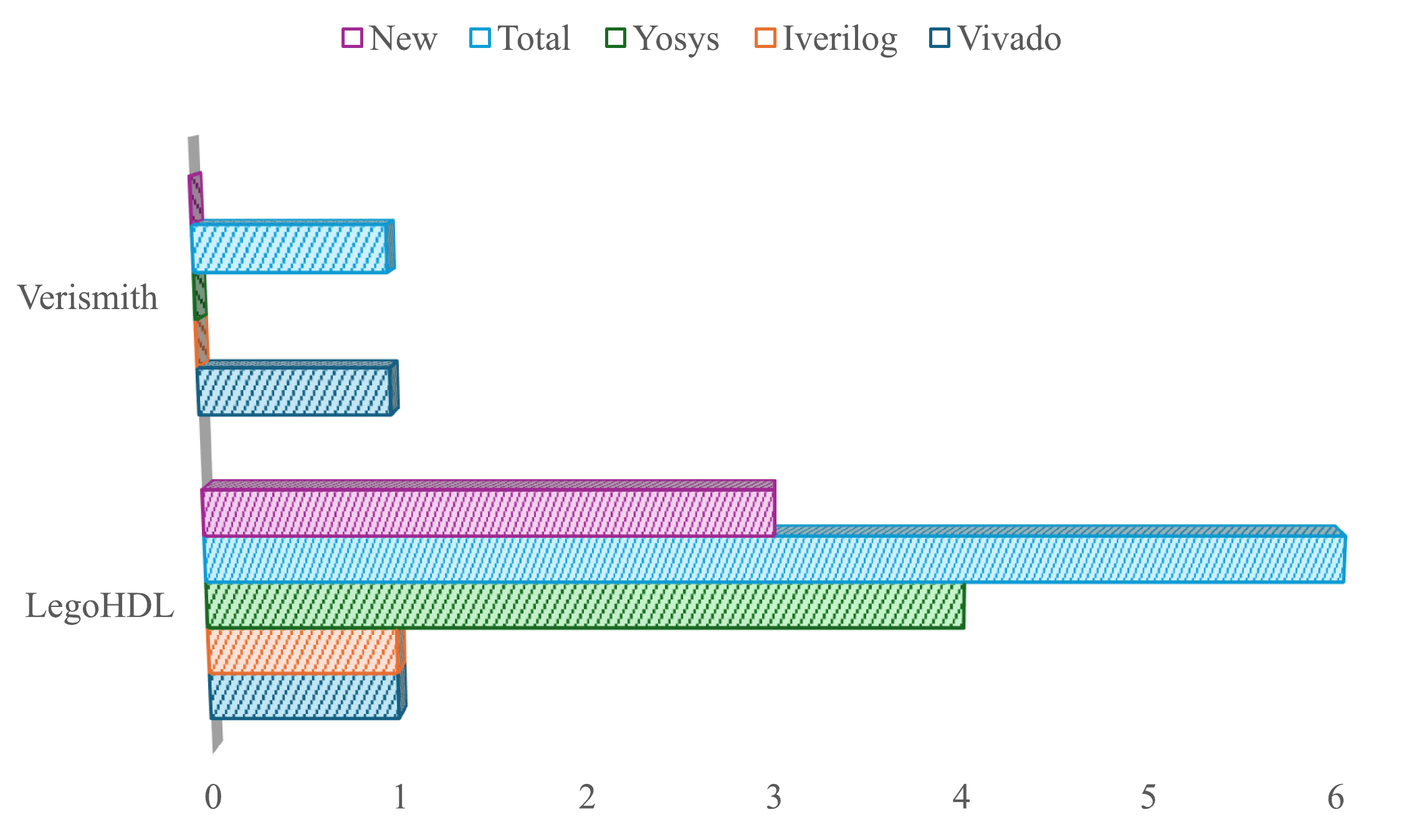}
\caption{Defects found by Verismith and LegoHDL}
\label{fig:8}
\end{figure}

\textbf{Result.} Table~\ref{exp2} and~\figurename \ref{fig:8} categorize the defects detected in our experiment into new defects (New) and previously identified defects (Known). 
It is evident that LegoHDL outperforms the state-of-the-art method, Verismith~\cite{verismith}, in identifying defects. 
Over a two-week period, LegoHDL identified six defects, of which three were new. 
In contrast, Verismith~\cite{verismith} detected only one defects, all of which were already known in the defect repository. 
Furthermore, the experiment revealed that while Verismith~\cite{verismith} possesses a specific capability in detecting defects within Vivado, it fails to identify defects in Yosys and Iverilog. However, LegoHDL demonstrates a more balanced capability in defect detection than Verismith~\cite{verismith}. 
This is attributed to LegoHDL's access to a more extensive corpus through the use of the Simulink HDL block library. 
By generating an AST to create CPS models and converting them into HDL code, LegoHDL also presents more complex data and control flows compared to Verismith~\cite{verismith}. These issues will be further explored in Section~\ref{diversity}.

\textbf{Conclusion.} LegoHDL can discover defect in FPGA logic synthesis compilers. And the capability of LegoHDL for finding defects is better than our SOTA.



\begin{table}[!t]
\caption{Defects found by Verismith and LegoHDL}
\begin{center}
\begin{threeparttable}
\begin{tabular}{cccccc}
\toprule
& \multirow{2}[2]{*}{Vivado} & \multirow{2}[2]{*}{IVerilog} & \multirow{2}[2]{*}{Yosys} & \multicolumn{2}{c}{Total}\\
        \cmidrule(r){5-6} 
& & & &New   & Known  \\
\midrule
Verismith& 1 & 0 & 0 & 1 & 0 \\
LegoHDL& \textbf{1} & \textbf{1} & \textbf{4} & \textbf{3} & \textbf{3}   \\ 
\bottomrule 
\end{tabular}
\end{threeparttable}
\label{exp2}
\end{center}
\end{table}

\subsection{Answer to RQ3: Can LegoHDL improve the complexity of HDL code compared to the state-of-the art methods?}\label{diversity}

\textbf{Approach.} We have three metrics to determine the distinctiveness between the SOTA and our methods. 
These three metrics are derived from the AST. 
The first indicator is the number of nodes, that is, the number of blocks. 
We will tally the number of blocks in both the file generated by our model and the SOTA, and then compare the counts. 
In simpler terms, if an HDL design has more blocks, it is considered more complex. 

The second metric is the count of lines, that is, the count of connections. 
A higher number of connections indicates a higher likelihood of encountering defects in the model. 
The third metric focuses on specific reference relationships. 
Complex reference relationships are more likely to expose defects. 
However, having ten or eleven files doesn't mean the SOTA method, which can create only one file, is less efficient. The final result is determined by the number of partitions. 
We use LegoHDL and Verismith~\cite{verismith} to separately generate 1000 HDL code files. Setting the code scale are 700-1000 lines(Our SOTA suggests that this scale of code is the most efficience scale for finding defects) and compare the code generate by different method following these metrics.

\begin{figure}[!t]
\centering
 \includegraphics[width=0.95\linewidth]{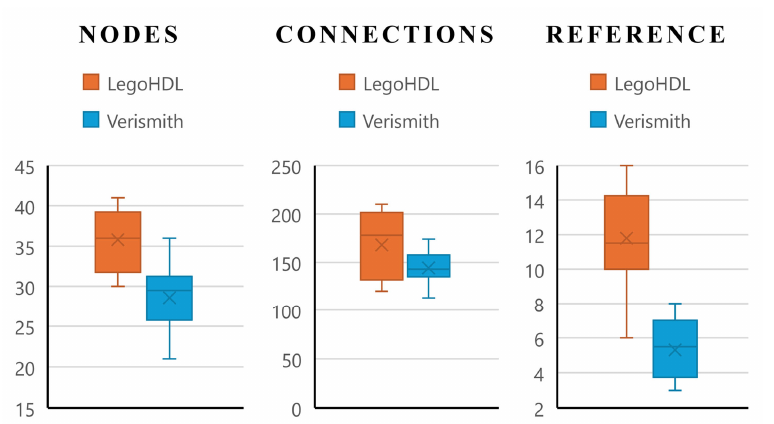}
\caption{Diversity comparison between LegoHDL and Verismith}
\label{fig:complexity}
\end{figure}

\textbf{Result.} \textit{Numbers of Nodes and Connections.} 
The most critical elements of HDL code are nodes and connections, essential for depicting the characteristics of HDL code. 
As illustrated in~\figurename \ref{fig:complexity}, the range of nodes and connections in HDL code generated by LegoHDL is [30,41] and [120,210]. It respectively exceeds those produced by Verismith~\cite{verismith}, which is [21,36] and [113,174]. 
Notably, LegoHDL demonstrates a significant advantage in terms of connections. In terms of the median number of nodes, LegoHDL is 31\% higher than Verismith, and in terms of the average number of nodes, LegoHDL is 25.1\% higher than Verismith. In terms of connection median, LegoHDL is 27\% higher than Verismith, and in terms of connection average, LegoHDL is 16\% higher than Verismith. These metrics indicate a higher complexity of connection relationships due to the AST-guided generation of CPS models. 
This suggests that LegoHDL can produce more complex HDL code in terms of data flow.

\textit{Specific Reference Relationships.} In HDL design, appropriate application relationships can optimize circuit design. 
While consolidating all code into a single file is possible, complex reference relationships pose a challenge to the compilation tool's execution capabilities. 
To quantify this metric, we fix the top-level design model and counted the number of model it references. \figurename \ref{fig:complexity} reveals that the range of model references generated by LegoHDL surpasses those by Verismith~\cite{verismith}, with LegoHDL outperforming Verismith~\cite{verismith} by 16.67\% regarding the median number of reference metrics.


\textbf{Conclusion.} LegoHDL is capable of generating more complex HDL code in both dataflow and control flow. This complexity contributes to the superior defect-finding capabilities of LegoHDL compared to Verismith in most instances.

\subsection{How efficient is LegoHDL in generating HDLCode?}

\textbf{Approach.} 
In this experiment, we aim to explore the relationship between the number of CPS models and the time of generated HDL code. This investigation will allow us to better control the size of generated HDL code and verify the efficiency of LegoHDL in generating HDL code. We divide the number of CPS models into 6 intervals, from 0 to 600. For each interval, we generate 100 CPS models, convert them into HDL code, and then record the time to generate HDL code.

\begin{figure}[!t]
\centering
 \includegraphics[width=0.95\linewidth]{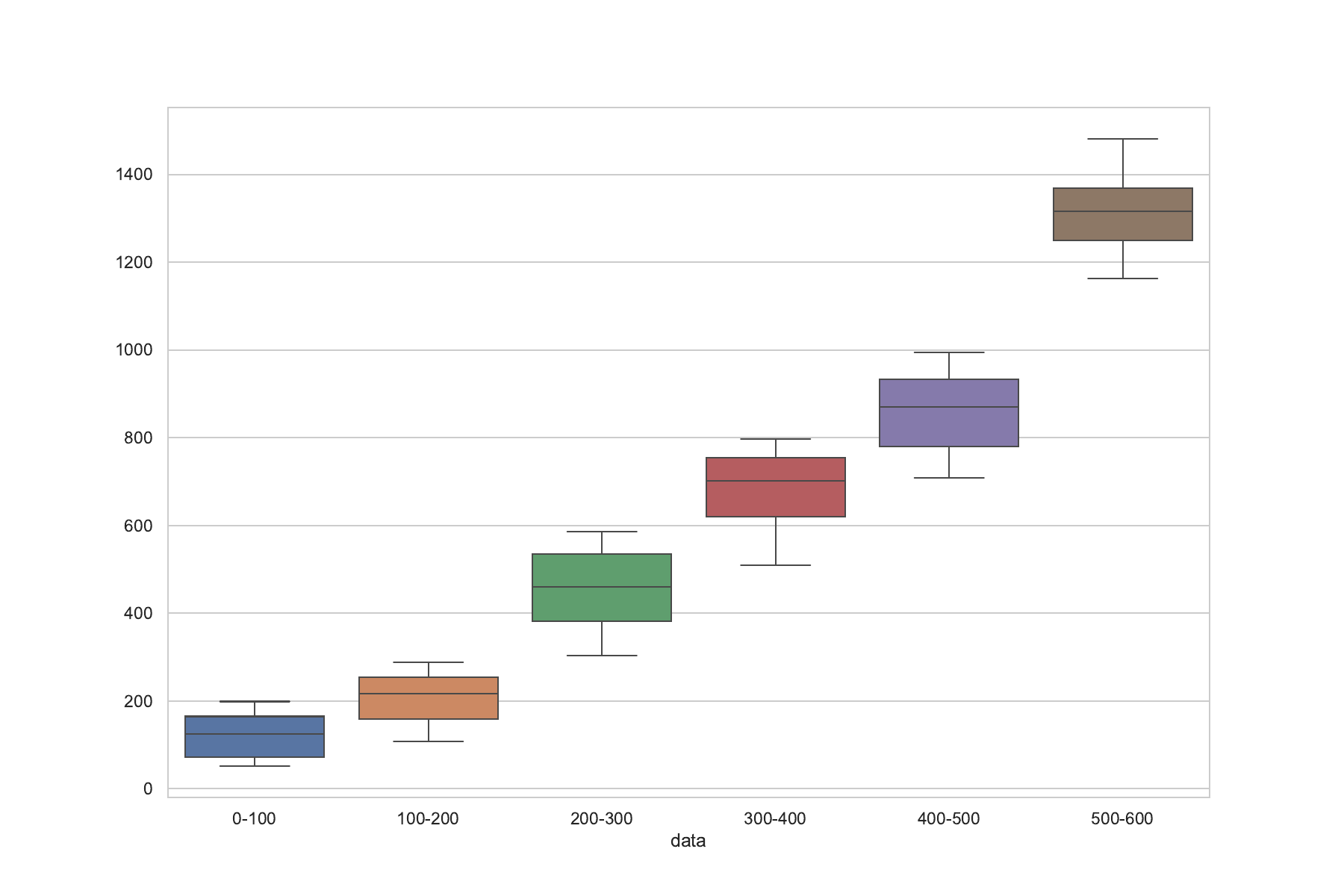}
\caption{Model size and generation time}
\label{fig:numbers}
\end{figure}

\textbf{Result.} The relationship of CPS block counts and scale of HDL Code has shown in~\figurename~\ref{fig:numbers}.
When the number of models ranges from 0 to 100, the translation time for the corresponding HDL code from CPS is approximately 50 to 200 seconds. As the model count increases to 200, the translation time for HDL code rises to about 300 seconds. Upon reaching 300 models, the time to generate HDL code varies from approximately 300 to 600 seconds. At 400 models, the generation time for HDL code exceeds 800 seconds. With 400 to 500 models, the HDL code generation time spans 700 to 1,000 seconds. Increasing the model count to 600 leads to a maximum HDL code generation time of 1,400 seconds. This escalation in time is attributed to the increasing complexity of functions and reference relationships as the number of models rises, consequently extending the translation time.

\textbf{Conclusion. }The time required for LegoHDL to generate HDL code will increase as the number of CPS models increases. However, the efficiency of code generated by LegoHDL generally falls within the expected parameters.

\section{Related Works}\label{related}

\subsubsection{FPGA Tool Chains Testing}

In the realm of FPGA synthesis compilers testing, Verismith~\cite{verismith} stands out as the primary methodology. 
It operates as a Verilog program generator, crafting random behavioral Verilog code without undefined values, based on predefined parameter configurations. 
Nevertheless, the capability of Verismith~\cite{verismith} to generate complex HDL code is somewhat limited as it exclusively produces Verilog, one type of hardware design language. 
To address these limitations and enhance the thoroughness of FPGA logic synthesis compilers testing, we introduced LegoHDL, aiming to ensure the accuracy and correctness of FPGA synthesis processes.

Another tool in this domain is VlogHammer~\cite{VlogHammer}, a Verilog fuzzer designed for testing major commercial FPGA logic synthesis compilers and several simulators. 
To date, VlogHammer~\cite{VlogHammer} has identified approximately 75 defects. 
However, unlike Verismith~\cite{verismith}, it does not generate multi-model programs and lacks support for behavioral-level Verilog constructs, such as \texttt{always} blocks. 
Similar to Verismith~\cite{verismith}, VlogHammer~\cite{VlogHammer} also restricts its output to Verilog.

Additionally, random Verilog generators like VERGEN~\cite{VEGEN} exist, which produce behavioral-level Verilog by randomly assembling high-level logic blocks, including state machines, MUXes, and shift registers. 
These generators, however, rely on predefined structures to construct their configurations, resulting in limited diversity. 
This limitation diminishes their capacity to test a broad spectrum of Verilog structure combinations. 
Furthermore, VERGEN~\cite{VEGEN} employs AFL~\cite{AFL}, a versatile fuzzer for binary files, leveraging instrumentation to guide test case mutations. 
Given the intricate nature of FPGA logic synthesis compilers and their myriad states, it may be challenging for fuzzers to pinpoint an effective strategy for test program generation. 

In recent years, Large Language Models (LLMs) have been widely used in test case generation. The same methods are employed in FPGA logic synthesis compilers testing. 
The most utilized method,  
 VeriGen~\cite{VeriGen}, employs a fine-tuned open-source CodeGen-16B model as a substitute for the commercial state-of-the-art GPT-3.5-turbo. 
It demonstrates a 41\% improvement in generating syntactically correct Verilog code across various problem categories compared to the pre-trained LLM generation method. 
However, while VeriGen excels in generating code, it is not specifically focused on testing; its capability in test case generation is limited by its corpus—a Verilog dataset compiled from GitHub and Verilog textbooks. 
This means the Verilog program generated by VeriGen may focus on specific aspects but cannot thoroughly test FPGA logic synthesis compilers. 
Furthermore, VeriGen cannot generate stimulus files, which further constrains the accuracy of testing. Other methods, such as BetterV~\cite{BetterV}, focus on optimizing Verilog code and reducing runtime. 
Currently, due to limitations in the accuracy of the corpus and the scale of model generation, LLMs are not sufficiently effective for HDL generation testing.

Moreover, numerous efforts have been dedicated to enhancing the reliability of FPGA development compilers. 
For instance, Yann Herklotz et al.~\cite{vericert} have focused on bolstering the stability of high-level synthesis (HLS) tools through formal verification, proposing Vericert, a formally verified HLS tool. 
Zewei Du et al.~\cite{zedu} have explored the application of fuzz testing to exhaustively evaluate HLS tools by supplying a vast array of valid C programs.

Despite these contributions, there remains a gap in the quality and complexity of the generated files, with a singular focus on one programming language. 
Consequently, we have embarked on a more comprehensive testing of FPGA logic synthesis compilers to affirm their stability and reliability.

\subsubsection{Differential Testing in Compiler Testing}

Differential testing~\cite{ChenHHXZ0X16,R31,R32,R50} considers the generation of equivalent variants of HDL code based on the input of a program. 
It detects defects by comparing the outcomes of different variants. 
Differential testing has been thoroughly validated in widely used compilation tools such as GCC and LLVM, identifying over a thousand defects~\cite{ChenHHXZ0X16,ChenPPXZHZ20,CaiL2019just,TangJZLRK22,TangRKJ20,JiangZRZL22,Hicond}.

In compiler testing, there are three EMI (Equivalent Moudle Input)-based differential testing mutation methods, including Orion~\cite{orion}, Athena~\cite{athena}, and Hermes~\cite{Hemers}. 
Orion~\cite{orion} focuses on mutating dynamic dead code regions by randomly pruning unexecuted statements to generate variant programs, while Athena~\cite{athena} is capable of inserting or removing code in these areas under different inputs.
Unlike Orion~\cite{orion} and Athena~\cite{athena}, which only mutate in dead code areas, Hermes~\cite{Hemers} can mutate both live and dead code areas to produce equivalent variants.

Beyond EMI-based differential testing, Jiang et al.~\cite{JiangZRZL22} introduced CTOS, which employs arbitrary optimization sequences to identify compiler defects in LLVM. 
Their method significantly enhanced the capability to detect defects. 
Tang et al.~\cite{R19} proposed a diversity-guided program mutation method to detect compiler warning defects. 
Chen et al.~\cite{R21} introduced a history-guided configuration diversification method for testing compilers.

\section{Conclusion and Future Work}

In this paper, we introduce LegoHDL to generate complex and diverse HDL code for logic synthesis compilers testing.
LegoHDL comprises two components: the function generation component and the syntax guidance component. 
The former component leverages extensive CPS model libraries to create diverse CPS function models.
The latter component uses HDLCoder to convert CPS function models into HDL code, and analyzes the corresponding AST for guiding the function generation component to insert additional CPS function models at a certain insertion position.
The two components continue to interact until the final HDL code and corresponding CPS model are generated, which is similar to Lego by piecing together small pieces of discrete functional component into a complete large project.
Within three months, the HDL code generated by LegoHDL has detected 20 defects -- many of which are deep and important in logic synthesis compilers.

In the future work, we plan to further improve LegoHDL by generating more complex HDL code. 
In addition, we plan to conduct an empirical study to deeply compare the effectiveness 
of different testing methods on more FPGA development tools.

\section*{Acknowledgment}

This work was supported by the National Natural Science Foundation of China (No.62202079, No.62032004),  the Dalian Excellent Young Project (2022RY35).

\bibliographystyle{IEEEtran}
\bibliography{GUIDANCE, IEEEabrv}
\vspace{12pt}

\end{document}